\documentclass{aastex}
\usepackage{emulateapj5}

\shorttitle{On the Detection of Sodium in HD 209458b}
\shortauthors{Fortney, et al.}

\newcommand{\hd}{HD 209458b}
\begin{document}

\slugcomment{\bf}
\slugcomment{Accepted to The Astrophysical Journal}

\title{On the Indirect Detection of Sodium in the Atmosphere of the 
Planetary Companion to HD 209458}

\author{J. J. Fortney\altaffilmark{1}, D. Sudarsky\altaffilmark{2}, I. Hubeny\altaffilmark{3}$^,$\altaffilmark{4}, C. S. Cooper\altaffilmark{1}, W. 
B. Hubbard\altaffilmark{1}, A. Burrows\altaffilmark{2}, J. I. 
Lunine\altaffilmark{1}} 

\altaffiltext{1}{Lunar and Planetary Laboratory, The University of 
Arizona, Tucson, AZ
\ 85721-0092;
jfortney@lpl.arizona.edu, curtis@lpl.arizona.edu, hubbard@lpl.arizona.edu, 
jlunine@lpl.arizona.edu}
\altaffiltext{2}{Department of Astronomy and Steward Observatory, The
University of Arizona, Tucson, AZ \ 85721; sudarsky@as.arizona.edu,
burrows@as.arizona.edu}
\altaffiltext{3}{NASA Goddard Space Flight Center, Greenbelt, MD 20771; 
hubeny@tlusty.gsfc.nasa.gov}
\altaffiltext{4}{NOAO, Tucson, AZ 85725}

%\newpage

\begin{abstract}

Using a self-consistent atmosphere code, we construct a new model of 
the atmosphere of the transiting extrasolar 
giant planet HD 209458b to investigate the disparity between
the observed strength of the sodium absorption feature at 589 nm and the 
predictions of previous models.  For the atmospheric temperature-pressure profile we derive, 
silicate and iron clouds reside at a pressure of several mbar 
in the planet's atmosphere.  These 
clouds have significant vertical extent and optical depth due to our slant viewing geometry
and 
lead to increased absorption in bands
directly adjacent to the sodium line core.  Using a non-LTE sodium ionization model
that includes photoionization by stellar UV flux, collisional processes with H$_2$, and radiative 
recombination, we show that the ionization depth in the
planet's atmosphere reaches $\sim$1/2 mbar at the day/night terminator.  
Ionization leads to a slight weakening of the sodium feature.  We present our baseline 
model, including ionization 
and clouds, which falls near the observational error bars.  The sensitivity of our conclusions to 
the derived atmospheric temperature-pressure profile is discussed.

\end{abstract}


\section{Introduction}

The planet in the system HD 209458 has been an object of intense study 
since
the first observations of its transit \citep{2000ApJ...529L..41H, 
2000ApJ...529L..45C}.  These
observations led
to a determination of the planet's radius and physical mass, confirming 
that it must be a gas giant. 
At wavelengths
where the atmospheric opacity is high, the planet blocks more
light, yielding a deeper transit light curve that makes the planet
appear physically larger.  The location and strength of these absorption 
features serve as diagnostics
of the
temperature and chemistry of the planet's atmosphere. \citet{2000ApJ...537..916S} predicted
 variations in the transmission spectrum of HD 209458b
due to neutral Na and K and to singly-ionized He. 
\citet{2001ApJ...553.1006B} explored effects such as ionization and 
winds and predicted large variations due to
H$_{2}$O, CO, and CH$_{4}$ in the infrared. 
\citet{2001ApJ...560..413H} explored a variety of physical effects that 
could
be found in transits, including refraction and the angular 
redistribution of photons due to
Rayleigh scattering.  These were found to be minimal for HD 209458b;
\citet{2001ApJ...560..413H} went on to derive the planet radius as a 
function of wavelength from 300 to
2500 nm.

Confirmation that absorption features due to the opacity of gaseous 
species could be
observed with
current technology was obtained by \citet{2001ApJ...552..699B}.  They 
observed the
transit of
HD 209458b from 581 to 638 nm with the STIS instrument on
\textit{HST} and obtained photometric accuracy near 100 
micro-magnitudes. 
This spectral region encompasses the Na-D doublet at 589 nm.  Using 
these data,
\citet{2002ApJ...568..377C} found that the transit was deeper by $(2.32 
\pm 0.57) \times 10^{-4}$ in
a narrow band
centered on the
star's Na-D lines than in adjoining bands at shorter and longer 
wavelengths. 
Importantly, the observed difference in transit depth in and out of the 
Na-D line is smaller than previously 
predicted
\citep{2002ApJ...568..377C}.  The nominal model of 
\citet{2001ApJ...560..413H} predicted a
difference between these two bands of $4.7 \times 10^{-4}$.

\citet{2002ApJ...568..377C}, using the models of 
\citet{2001ApJ...553.1006B},
performed a parameter study to investigate a variety of possible reasons 
for this difference.  These
included a 
bulk underabundance of Na, the sequestering of atomic Na in condensates 
and/or
molecules, clouds very high in the planet's atmosphere,
and photoionization of atomic Na due to the UV flux from the parent 
star.  
\citet{2002ApJ...569L..51B} suggested that Na is not in local 
thermodynamic
equilibrium (LTE).  They speculated this leads to a core reversal of the 
589-nm absorption doublet in the planet's emission
spectrum, but they did not calculate its effect on
a transit light curve. 

Here we explore the
effects of cloud formation and ionization of Na in HD 209458b's 
atmosphere.  We find that cloud opacity is the dominant effect that determines the depth of the Na-D feature, but 
ionization of 
Na also leads to a non-negligible weakening of the Na-D doublet.

In our analysis, following \citet{2002ApJ...568..377C}, we define a narrow ``in feature" wavelength 
band as the 
wavelength 
range 
588.7-589.9 nm, and an ``out of feature" band as the combination of the 518.8-588.7 and 
589.9-596.8 
nm wavelength ranges.  
The ``in feature" band includes the wavelength extent of the easy-to-observe HD 209458A 
(stellar) Na absorption doublet.  When quoting a depth for the Na-D feature, we simply mean the
difference in transit depth (the fractional change in flux) between the ``in feature" and ``out 
of feature" bands at midtransit.  The fractional change in flux at midtransit across the entire 
581 to 638 nm wavelength range is -0.0164, meaning 1.64\% of the star's light is blocked.

\section{Temperature-Pressure Profile}
The temperature-pressure (\emph{T-P}) profile for a close-in extrasolar 
giant planet is flatter (more isothermal) than that
of an isolated object with the same effective temperature because of
heating of the outer atmosphere by the parent star.  In the present work, 
we have employed TLUSTY \citep{HubenyLanz}, a self-consistent atmosphere
code that uses the Discontinuous Finite Element (DFE) method 
for radiative transfer \citep{HubenyLanz,Castor92}.  The atmosphere model is
plane parallel.  
Radiative equilibrium is achieved by adjusting the
atmospheric temperature-density structure iteratively via
the Accelerated Lambda Iteration (ALI) method.  The important physical
quantities, including net flux, temperature, and density
in every atmospheric depth zone, are converged to better than 0.1\%,
and total energy is conserved to this same precision.  For the deep convective
layers a mixing-length prescription is used with a mixing-length parameter of
one scale height.

For a close-in planet such as HD 209458b, an inner boundary flux
corresponding to that of an isolated object with an effective
temperature of $\sim$500 K or less results in an outer atmospheric
structure that is independent of this inner boundary. The outer
structure is determined by the stellar flux, the surface gravity, the atmospheric compositions,
and the opacities \citep[see][]{Sudar02,2001RvMP...73..719B}.  
The metallicity of the parent star has been shown to be consistent with solar abundances
\citep{2000ApJ...532L..55M}, and we use these abundances when modeling the planet's atmosphere.

The \emph{T-P} profile of HD 209458b's atmosphere that we derive is
shown in Fig.~\ref{figure:pt}.  We use a surface gravity of 980 
cm/s$^2$ for the
planet. 
We find that HD 209458b is a ``Class V" 
planet as defined in
\citet{2000ApJ...538..885S} and \citet{Sudar02}, meaning that silicate and iron clouds occur 
at altitudes in the
planet's atmosphere relevant to spectroscopic observations.  We model 
the location, distribution, and size of cloud particles using the
model of \citet{2002astro.ph..5192C}.  As seen in Fig.~\ref{figure:pt}, 
the cloud bases of forsterite (Mg$_2$SiO$_4$) and
iron (Fe) are at the several mbar level.  The effect of these clouds on the transit 
is discussed in Section 4.

In calculating the \emph{T-P} profile, we assume that incoming stellar 
energy is absorbed and redistributed evenly on the day side, but is not 
redistributed onto the night side.  In a recent paper, 
\citet{2002A&A...385..166S} find that
redistribution of energy around the entire planet may be inefficient, 
leading to a $\sim$500 K
effective temperature difference between the planet's day and night 
sides.  The transit observation 
is sensitive to
the planet's atmosphere on the limb, which could have a \emph{T-P} 
profile
slightly cooler than we predict here.  (Note that the atmospheric 
circulation
code of \citet{2002A&A...385..166S} is not able to calculate \emph{T-P} 
profiles for the planet.)  Based on
this possibly large temperature difference,
\citet{2002A&A...385..156G} make the point that if HD 209458b does have 
a
500 K effective temperature contrast between the night and day sides, 
condensates could form on the night side and tie up Na, thereby
weakening the Na-D absorption feature. 
Note, however, that the ``cold" \emph{T-P} profile used in 
\citet{2001ApJ...560..413H}, with a
temperature 400 K cooler at 1 mbar than seen in Fig.~\ref{figure:pt}, 
still
has a 589-nm Na feature depth of $3.2 \times 10^{-4}$, deeper than 
observed.

Our \emph{T-P} profile differs significantly from that calculated in the 
recent paper on HD 209458b
by \citet{2002ApJ...569L..51B}.  We find a temperature of 1280 K at 1 
mbar and 950 K at 10 $\mu$bar. 
Examination of Fig.~1 in \citet{2002ApJ...569L..51B} shows temperatures
of $\sim$1790 K at 1 mbar and 1630 K at 10 $\mu$bar.  The reasons for these large differences are unclear, but could be due to differing opacities, stellar flux, or planet surface gravity.  Our T-P profile is 
closer to those for \hd\ discussed by \citet{2000ApJ...537..916S} and 
\citet{2002astro.ph.10612R}.  These
differences are significant,
and we are led to very different conclusions concerning the physical 
processes of importance for the Na-D transit depth.  As with late L and 
T dwarfs, our HD 209459b atmosphere is free of metals such as Mg, Al, Ca, 
Fe, and Si, which have been
incorporated into condensates.  \citet{2002ApJ...569L..51B} find
an atmosphere that is not cool enough to allow for the formation of 
these condensates, which leaves
their
atmosphere rich in heavy metals in atomic form. Due to the high UV 
opacity of free
metals that remain in the atmosphere and absorb stellar radiation, 
\citet{2002ApJ...569L..51B}
conclude 
that ionization of Na at pressures greater than 1 $\mu$bar is
negligible. Our model temperatures allow deeper ionization of Na in the atmosphere of the planet. 

\section{Transit Calculations}

The methods we use for calculating the effects of absorption by gaseous
species, absorption and scattering by condensates, Rayleigh scattering, 
and refraction are
essentially identical to those described in \citet{2001ApJ...560..413H}, 
although we remove the simplification that the opacity versus pressure profile 
is the 
same around the entire planet.

We calculate the slant optical depth through the planet's atmosphere,
$\tau$, from the expression
\begin{equation} \label{eq:Tau}
\tau=\int_{-\infty}^{r}\frac{r^{\prime\prime}dr^{\prime\prime}\sigma(r^{\prime\prime}) N(r^{\prime\prime})}
{{(r^{\prime\prime2}-r^{2})}^{1/2}} + \int_{r}^{\infty}\frac{r^{\prime\prime}dr^{\prime\prime}\sigma(r^{\prime\prime}) N(r^{\prime\prime})}
{{(r^{\prime\prime2}-r^{2})}^{1/2}},
\end{equation}
where $\it{r}$ is the impact parameter of a light ray, $r^{\prime\prime}$ is the magnitude of the 
three-dimensional vector position of an atmospheric point from the planet's center, $\sigma$
is the absorption cross-section per molecule, and $\it{N}$ is the number density
 of molecules in the
atmosphere.  Two terms in the $\tau$ expression are needed because $\sigma(r^{\prime\prime})$ is 
different on the day and night sides.

As before, we use the planetary and stellar properties from 
\citet{2001ApJ...552..699B}. 
Fig.~\ref{figure:pvsl} shows, for a cloudless atmosphere with no 
ionization, the
radius as a function of wavelength over the range covered by the 
\citet{2001ApJ...552..699B} and \citet{2002ApJ...568..377C}
observations.  The auxiliary axis of Fig.~\ref{figure:pvsl} shows the pressure at the terminator 
to which this radius
corresponds; 
the figure thus indicates the pressure range probed by the transit 
of HD 209458b.

\section{Cloud Model}
\subsection{Location, Distribution, and Opacity}
The \emph{T-P} profile we derive is warmer than the \emph{ad hoc} profiles studied 
in 
\citet{2001ApJ...560..413H}.  Because of this, we find forsterite and 
iron clouds higher in the planet's
atmosphere.  If we examine the forsterite cloud in Fig.~\ref{figure:pt}, for the \emph{T-P} 
profile 
without cloud opacity, the cloud base is at 5.4 mbar.  The opacity due to clouds slightly warms the 
planet's atmosphere, especially at the location of the clouds, moving the cloud base farther out, to 
1.8 mbar.  Predicting the exact location of the cloud base is a non-trivial problem, and in a fully 
self-consistent calculation of cloud location, one would need to iterate many times successively 
between the 
cloud code and atmosphere code.  Since cloud opacity is a strong function
of particle size, and the predicted particle sizes are altered due to a 
change in the cloud base and as function of pressure $within$ the cloud 
itself (as described below), this step is challenging.  Since 
uncertainties remain in the modeling of clouds, and the effects of 
redistribution of energy on the \emph{T-P}, we do not attempt this fully 
self-consistent treatment.  Instead, our \emph{T-P} profile shown in 
Fig.~\ref{figure:pt} incorporates the opacity of the cloud with the 
location of the base given by the point of intersection with 
the condensation 
curve, and with a cloud particle size given by the modal size for the 
entire cloud.  The particle number density falls of exponentially for one 
scale height.

The cloud model of 
\citet{2002astro.ph..5192C} predicts particle radii in this radiative 
region for the iron cloud that range from tens of $\mu$m at the base 
to less than a $\mu$m at the cloud tops.  In the 
forsterite cloud, all particles are 0.05 to 0.5 $\mu$m in size.  The 
cross-sections are calculated as in \citet{Sudar02}.  For these submicron particles, 
scattering dominates over absorption, and the opacity of the forsterite cloud is significantly larger 
than that of the iron cloud.  The \citet{2002astro.ph..5192C} model predicts a 
distribution of cloud particle sizes and densities as a function of pressure in the atmosphere. 
Therefore, we calculate the cross-sections for the cloud as a function of pressure at many 
pressures 
within the cloud and put no artificial cap on the cloud top.  While understanding the vertical distribution of cloud 
material 
is a well-known area of importance for brown dwarf atmospheres, it is of even greater 
importance for the transit spectroscopy of EGP atmospheres, because cloud opacity can be so 
great for our slant viewing geometry.

For \hd, we find a forsterite cloud with 
significant vertical extent, with a large slant optical depth ($\tau$ $\sim$500 at the base) 
that does not fall to $\tau$ $\sim2/3$ until 1.9 gas scale heights above the cloud base.  Even 
though 99\% of the 
cloud's mass is found within 1 gas scale height, a non-negligible population of small, highly 
scattering cloud particles continues above this level, giving rise to a forsterite cloud with a 
significant
\emph{slant} optical depth up to $\sim$1.9 
scale heights.  Since our predicted forsterite cloud is in a region of interest for \hd's transit 
spectroscopy, the vertical extent of the cloud is of great importance.  It seems likely that any 
clouds seen during transit spectroscopy will be in a radiative (quiescent) region, in which case the cloud 
particles will likely remain small, in contrast to the large sizes (10-200 $\mu$m) predicted for convective regions.  Given current 
uncertainties in cloud modeling \citep[see][]{2002astro.ph..5192C}, much work still needs to be 
done in this area.  While \citet{2000ApJ...537..916S} were the first to note the importance of 
clouds in determining the depth of transit absorption features, here we note that understanding the 
vertical distribution of clouds is yet a further challenge in EGP transit spectroscopy.

In this narrow spectral 
region (581-638 nm), the cross-sections have virtually
no wavelength dependence.  Thus, clouds lead to a wavelength-independent 
absorption of stellar light greater than that of the gaseous species
alone, leading to a uniform increase in the planetary radius across much of 
the wavelength band.  In the core of the Na-D feature, where the slant
optical depth reaches unity at much lower pressures, the clouds have no 
effect.  This leads to a decrease of the contrast in transit depth
between the ``in feature" and ``out of 
feature" bands.  Fig.~\ref{figure:pvsl2} is analogous to Fig.~\ref{figure:pvsl}, but in this case 
shows the planet's radius and pressure at the terminator as a function of wavelength, for our 
atmosphere including clouds, with the base at 5.4 mbar.  In both cases, the planet's radius at 1 
bar has been adjusted so that 
the transit depth remains 1.64\%.

Fig.~\ref{figure:dvsl2} shows the transit depth versus wavelength for a 
neutral cloud-free atmosphere and a neutral atmosphere with the 
predicted clouds.  The cloud-free model gives a 
feature depth
of $5.20 \times 10^{-4}$, deeper than that obtained by 
\citet{2001ApJ...560..413H} because of our warmer \emph{T-P} profile.  The 
model with clouds decreases this 
depth significantly.  With the forsterite cloud base at 5.4 mbar, the depth is $2.42 \times 
10^{-4}$, while if the cloud base is at 1.8 mbar the depth is $0.93 \times 10^{-4}$.  The depth is 
actually $smaller$ than the \citet{2002ApJ...568..377C} observations, $(2.32 
\pm 0.57) \times 10^{-4}$, indicating that if clouds form 
too high in the planet's atmosphere they can dominate over the opacity of gaseous species.

\subsection{Sensitivity of Results}
Fig.~\ref{figure:pt}, which shows the \emph{T-P} profile, along with the pressure axes on Figs. 
\ref{figure:pvsl} and 
\ref{figure:pvsl2}, highlight how sensitively the Na-D feature depth depends on the location of 
clouds.  
Moving the cloud base $\sim$1 gas scale height can cause an appreciable difference in the Na-D 
feature strength.  As Fig.~\ref{figure:pt} and the transit depths for the two models with clouds 
show, a change in the pressure of the cloud base of this amount would happen with fairly minor 
deviations 
from our derived \emph{T-P} profile for the planet's atmosphere.  If the cloud base occurs at 1.8 mbar, rather than 5.4 mbar, the feature depth moves from comfortably within the error bars to 
significantly below.  Even higher clouds (warmer \emph{T-P} profiles) would lead to even 
weaker Na-D features.   At several hundred K warmer, no condensate clouds would 
form, as is the case in the model of \citet{2002ApJ...569L..51B}.  Perhaps the most important 
issue is that changing 
our assumption of how absorbed stellar radiation is redistributed around the planet causes 
non-negligible changes to the planet's \emph{T-P} profile \citep[e.g.,][]{Sudar02}. 

We turn next to the role of ionization, and treat the case where the 
opacity profile varies over the
planet. We find that the depth to which the Na is ionized changes with 
angle from the
planet's
subsolar point.  In \citet{2001ApJ...560..413H}, Na ionization by UV 
irradiation was ignored.  Here we find that ionization of Na probably has a secondary but 
non-negligible effect on the observed Na-D feature.

\section{A Simple non-LTE Sodium Ionization Model}

The proximity to the parent star subjects the transiting planet to large 
amounts of ionizing radiation, making equilibrium chemistry treatments 
inadequate. 
In particular, Na ionization alters the profile of neutral Na.  This is 
crucial,
since the Na-D line and its transit signature are features of the 
neutral state.  A rigorous study of the ionization of sodium in the atmosphere
of \hd\ would require sophisticated non-LTE and non-equilibrium
photochemical modeling. However, most of the necessary atomic 
and molecular data (in particular, collisional rates and dissociation 
parameters) are either poorly known or completely unknown. 
Therefore, we attempt to calculate the effects of ionization using 
simple, parametric models.

\citet{2002ApJ...569L..51B} have performed a detailed study of non-LTE effects
on excitation and ionization of Na I in the atmosphere of \hd.
However, they find that the UV opacity is dominated by atomic metals 
(Mg, Al, Ca, Fe, and Si). Because of the much cooler temperatures we 
find, these elements are tied up in condensates and are thus
depleted from the outer layers of the atmosphere of \hd. Therefore, they 
cannot shield sodium from intense stellar ionizing radiation.  Moreover, 
\citet{2002ApJ...569L..51B} had to parameterize the corresponding 
collisional rates, and they found that free electrons are only due to 
ionization of potassium.

In view of all uncertainties in the atomic and molecular
parameters, and especially the UV opacities, we feel that it is not 
warranted to perform a detailed
non-LTE study. Instead, we devise a simple, non-LTE analytical model 
that allows us to study the effects of various uncertainties in relevant
quantities (opacities, free electron densities, collisional rates) on the 
degree of ionization of sodium in a straightforward  
way.

We find that much 
of the outer layers of the atmosphere (P $\lesssim$1/2 mbar), where Na I 
has the greatest likelihood of having non-LTE 
level populations, are ionized by the stellar UV flux.  We also calculate 
the change in the ionization depth as a function of angle from the 
substellar point, which to this point has not been investigated in the 
literature.

\subsection{Description of the Model}

We assume both neutral and ionized sodium are represented by one-level
atoms, with populations (number densities) $n_1$ and $n_2$.
The third unknown of the problem is the electron density, $n_{\rm e}$. 
The three equations
that specify these unknown number densities are the particle
conservation equation,
\begin{equation}
\label{pc1}
n_1 + n_2 = N_{\rm Na}\, ,
\end{equation}
charge conservation equation,
\begin{equation}
\label{chc1}
n_{\rm e} = n_2 + \alpha N_{\rm Na} + \beta n_2\, ,
\end{equation}
and the statistical equilibrium equation for sodium,
\begin{equation}
\label{se1}
n_1 [R_{12} + n_{\rm e}\, C_{12} + n_{\rm H2}\, C_{12}\, e_{\rm H2} ] =
n_2 [R_{21} + n_{\rm e}\, C_{21} + n_{\rm H2}\, C_{21}\, e_{\rm H2} ]\, ,
\end{equation}
where $N_{\rm Na}$ is the total sodium number density. In Eq.~(\ref{chc1}), we assume that the total number of electrons is given
by the total number of ionized sodium atoms, plus a contribution
from an unspecified source that is given through some fraction
$\alpha$ of the total number of sodium atoms (that is, it scales
as the total density), plus another empirical contribution that scales
as ionized sodium, with a multiplicative factor $\beta$. 
Setting $\alpha=\beta=0$ stipulates that free electrons are provided
only by ionization of sodium.

In the statistical equilibrium equation (\ref{se1}), $R_{12}$ and
$R_{21}$ are the photoionization and radiative recombination rates,
respectively; $ n_{\rm e} C_{12}$ and $n_{\rm e} C_{21}$ are the collisional
ionization and recombination (three-body recombination) rates with free
electrons. Since the dominant species in the atmosphere is H$_2$,
we have to consider ionization of sodium by collisions with
H$_2$ and its inverse process as well. In the absence of any
knowledge of relevant rates, we assume, similar to \citet{2002ApJ...569L..51B},
that the collisional rate is given by a scaled electron rate
with some factor $e_{\rm H2}$. It is reasonable to consider this
factor to be the ratio of reduced masses of the Na + H$_2$ pair
to that of Na + electron, which is about $1/60$. However, we keep
$e_{\rm H2}$ as a free parameter, allowing us to study the effects of the H$_2$ collision 
efficiency on Na ionization.

By detailed balancing arguments \citep[e.g.,][]{ 1978stat.book.....M}, the inverse 
rates are given by
\begin{equation}
\label{invrate}
C_{21} = C_{12}\, (n_1/n_2)^\ast = n_{\rm e}\, \phi(T)\, C_{12}\, ,
\end{equation}
where $(n_1/n_2)^\ast$ denotes the LTE value, and $\phi$ is the
Saha-Boltzmann factor, given by
\begin{equation}
\label{sb}
\phi(T)= 2.0706\times 10^{-16} \,(g_1/g_2)\, T^{-3/2} \exp(h\nu_0/kT)\, ,
\end{equation}
where $T$ is the temperature, $g_1$ and $g_2$ the statistical weights
($g_1=2, g_2 =1 $ for sodium), $\nu_0$ the ionization frequency,
and $h$ and $k$ the Planck and Boltzmann constants.

The radiative rates are given by
\begin{equation}
R_{12} = {4\pi\over h}\int_{\nu_0}^\infty {\sigma(\nu)\over\nu}
J_\nu d\nu\, ,
\end{equation}
and
\begin{equation}
\label{r21}
R_{21} = n_{\rm e}\, \phi(T)\, {4\pi\over h}\int_{\nu_0}^\infty {\sigma(\nu)\over\nu}
\left( {2h\nu^3\over c^2} + J_\nu \right) \exp(-h\nu/kT) d\nu\, ,
\end{equation}
where $\sigma(\nu)$ is the photoionization cross-section, and
$J_\nu$ is the mean intensity of radiation.  However, Eq.~(\ref{r21}) takes into account only 
recombinations
to the ground state of Na I. In reality, recombinations can occur to any state, and 
at low temperatures these processes are very
efficient. We therefore take the recombination rate to be
\begin{equation}
\label{c21vf}
C_{21}=5.461 \times 10^{-12} /\Big[\sqrt{T/T_0} \big(1+ \sqrt{T/T_0}\big)^{0.825}\Big]\, ,
\end{equation}
as given by \citet{1996ApJS..103..467V}.

The radiation intensity should in principle be
determined by solving the radiative transfer equation. We could
do so, but in view of gross uncertainties in relevant UV opacities,
this is not warranted. Instead, we simply assume that the specific intensity
is given as a sum of attenuated stellar irradiation intensity plus
a thermal component given by the Planck function at the local 
temperature. (This part is negligible at surface layers of interest, but
gives the correct behavior of intensity at depth.) 
To allow for uncertainties in UV opacities, we introduce another 
multiplicative factor, $w_{\rm add}$, which represents an additional 
attenuation of radiation, perhaps by an unspecified opacity source.

The intensity as a function of depth is given by
\begin{equation}
I_{\nu\mu}(\tau_\nu) = w_{\rm add}\, I_{\nu\mu}^{\rm inc} 
\exp(-\tau_\nu\mu) + B_\nu(T)\, ,
\end{equation}
where $\mu$ is the cosine of the angle with respect to the surface 
normal, $\tau_\nu(P)$ is the optical depth (in the normal direction), 
and $I_{\nu\mu}^{\rm inc}$ is the specific intensity 
of the stellar radiation at the top of the planetary atmosphere. 
We use a Kurucz model for a G0V star with a dilution factor of
$w = 1.4 \times 10^{-2}$ due to the planet-star distance.
For simplicity, we assume isotropic irradiation, which gives
\begin{equation}
\label{mean}
J_\nu \equiv {1\over 2} \int_{-1}^1 I_{\nu\mu} d \mu =
{1\over 2}\, w_{\rm add}\, I_\nu^{\rm inc} E_2(\tau)+ B_\nu(T)\, ,
\end{equation}
where $E_2$ is the second exponential integral.

The electron collisional ionization rate is assumed to be given by
Seaton's formula,
\begin{equation}
C_{12} = 1.55 \times 10^{13}\, T^{-1/2} \, \bar g\, \sigma(\nu_0)\,
\exp(-u_0)/u_0\, ,
\end{equation}
where $\sigma(\nu_0)$ is the photoionization cross-section at the edge,
$\bar g = 0.1$, and $u_0=h\nu_0/kT$.

\subsection{UV Opacities}

Since we find no free heavy elements in atomic form above the clouds in which these elements are 
sequestered, any opacity that may affect Na ionization must be 
provided by molecular species.  We have searched the literature,
and found only two equilibrium species that contribute
significantly to the opacity in the near UV---H$_2$O \citep{Yoshino96}, and H$_2$S \citep{Lee87}. 
Rayleigh scattering
on H$_2$ is much less important at this range. For both H$_2$O and H$_2$S data are only avaivable 
at 295 K. We plot in Fig.~\ref{figure:opac}
the total opacity, plus the individual contributions
of H$_2$O, H$_2$S, H$_2$, and a maximum possible contribution of
Na I (that is, assuming that sodium is completely neutral) at
upper layers of the atmosphere of \hd, as they follow from the atmosphere model.
The corresponding Na I photoionization cross-section is taken
from the Opacity Project database \citep{TOPBASE}.

\subsection{Representative Results}

The basic set of equations (\ref{pc1}) - (\ref{se1}), together
with auxiliary expressions (\ref{invrate}) - (\ref{mean}),
can be cast into a single cubic equation for $n_2/N_{\rm Na}$.
All the necessary 
quantities, $T$, $N_{\rm Na}$, $n({\rm H}_2)$, 
$n($H$_2$O), and $n(H$$_2$S), as functions of depth, are provided
by our model atmosphere for \hd.

Our nominal model, without any empirical modifications,
uses $\alpha=\beta=0$, $E_{\rm H2} = 1/60$, and $w_{\rm add} =1$.  The depth dependence of Na 
ionization for this model is shown in Fig.~\ref{figure:shield}.  This calculation corresponds to 
the planet's \emph{subsolar point}.  For the nominal model, the pressure at which the number
densities of Na I and Na$^+$ are equal is 1.6 mbar.  At greater pressures, neutral Na dominates.  
Deep in the planet's atmosphere, at pressures $> 100$ bar, ionization of Na again becomes 
appreciable due to the higher temperatures.  Also shown is the ionization depth dependence without 
shielding due to the other UV absorbers.  The pressure at which Na is 50\% ionized increases to 3 mbar.

We next test our predictions for Na ionization depth by varying several of the parameters 
mentioned earlier.  These trials can be seen in Fig.~\ref{figure:abw}.  In the first case $\alpha=1$, which represents the case where there is another donor,
 with a total abundance equal to that of sodium, which is fully ionized.  The model with $\beta=1$
represents an additional empirical donor of free electrons that
has the same abundance as {\em ionized} sodium.  The model with $w_{\rm add}=0.1$ represents the 
situation where unknown UV absorbers leave only 10\% of the calculated incoming ionizing flux 
available to ionize Na.  The electron donor factors seem like reasonable upper limits because 
species more abundant than Na in the planet's atmosphere tend to have 2-3 times larger 
ionization energies and consequently are difficult to ionize.  These factors have a modest 
effect on the pressure at which Na is 50\% ionized.  Changing the value of $e_{\rm H2}$ (which 
characterizes the H$_2$ collision efficiency) by factors of even 100 had negligible effects on the 
ionization depth for pressures $< 100$ mbar.

\subsection{Ionization on the Planet's Limb}
The important drawback of the ionization model outlined above is that the calculations are
representative of the subsolar point $only$.  We are interested in ionization at the planet's 
limb.  At a given location in the planet's atmosphere an angle $\theta$ from the subsolar point, the absorbed radiation of a given 
surface area will be diluted by a factor $\cos \theta$, meaning the ionization depth will not be as 
great.  Even this geometric view is too simple, as it would imply no photoionization at the 
planet's terminator.  To overcome this shortcoming, we have employed a somewhat different 
ionization model that can predict the ionization depth vs. angle around the planet.

In this angle-dependent 2-D ionization model (which we will call model 2, while the prior model is 
model 1), 
we assume incoming photons with energy greater than 5.14 eV (the ionization energy
of Na) and less than
11.2 eV, the energy at which significant H$_2$ absorption begins, ionize
atomic Na.  Na$^+$  is the only ionized species and
consequently the number of free electrons is equal to the number of   
Na$^+$ ions.
We use 2- and 3-body radiative recombination coefficients from   
\citet{1996ApJS..103..467V} and \citet{Gudzenko}.  Incoming photons
at some
oblique angle relative to the surface can ionize all the Na along a chord until, at some pressure, 
most ionizing photons (per second) have been used up, and recombinations (per second) can match 
ionizations (per 
second).  This is the ionization depth.   
Much as for a Str\"omgren sphere, we find a sharp
boundary between regions of the planet's atmosphere with complete
ionization and with
no ionization.  (Calculations from model 1 find a more gradual onset of neutral Na with depth.)  
We calculate the depth to which the atmosphere is ionized at several
hundred angular positions
from the planet's substellar point.  We find that Na is ionized to  
a depth of $\sim$12 mbar at the planet's substellar point (this was 1.6 mbar for model 1 with shielding due to UV absorbers, and 3 mbar without), and as 
one
moves away from this point, the pressure to which Na is ionized
decreases.  At the terminator (90 degrees from the subsolar point) the ionization depth
is 1.7 mbar.  Fig.~\ref{figure:pvsa}  
shows representative calculations for the change in ionization depth   
with
angle near the terminator.  At an angle of $\sim$10 degrees past the terminator, no Na is
ionized.  The model possesses cylindrical symmetry about
the planet's subsolar-antisolar axis.  The removal of the opacity
of atomic Na in the 10 mbar to 1 $\mu$bar region of the
atmosphere produces a negligible difference in the \emph{T-P} profile.

The column density of atmosphere through which photons travel until one reaches the ionization 
depth of Na can be calculated at the subsolar point for model 1, and at 
any angle, including the subsolar point for model 2.  Using the ratio of 
the column depths of the two models at the subsolar point, one can scale the results from model 2 to calculate the ionization depth around the planet for model 1.  Fig.~\ref{figure:pvsa} shows the ionization depth versus angle for the scaled model 
1 results.

Fig.~\ref{figure:dvsl} depicts how ionization leads to a shallower 
transit depth.  The differences between the \citet{2002ApJ...568..377C} 
``in feature" and ``out of feature" bands are, for model 1,  $4.72 \times 10^{-4}$ and for model 2, 
$4.83
\times 10^{-4}$.  Neither of these models include the predicted clouds, so that 
the relatively minor effect due to ionization can be shown clearly.  
Fig.~\ref{figure:pvsa} shows a model in  
which 
we increased the pressure level to which Na is ionized, in order to gauge 
how much ionization is needed to match the \citet{2002ApJ...568..377C} 
observations without the opacity of clouds.  The feature depth for this 
model is $2.87 \times 10^{-4}$.

\section{Conclusions}

\begin{figure}
\plotone{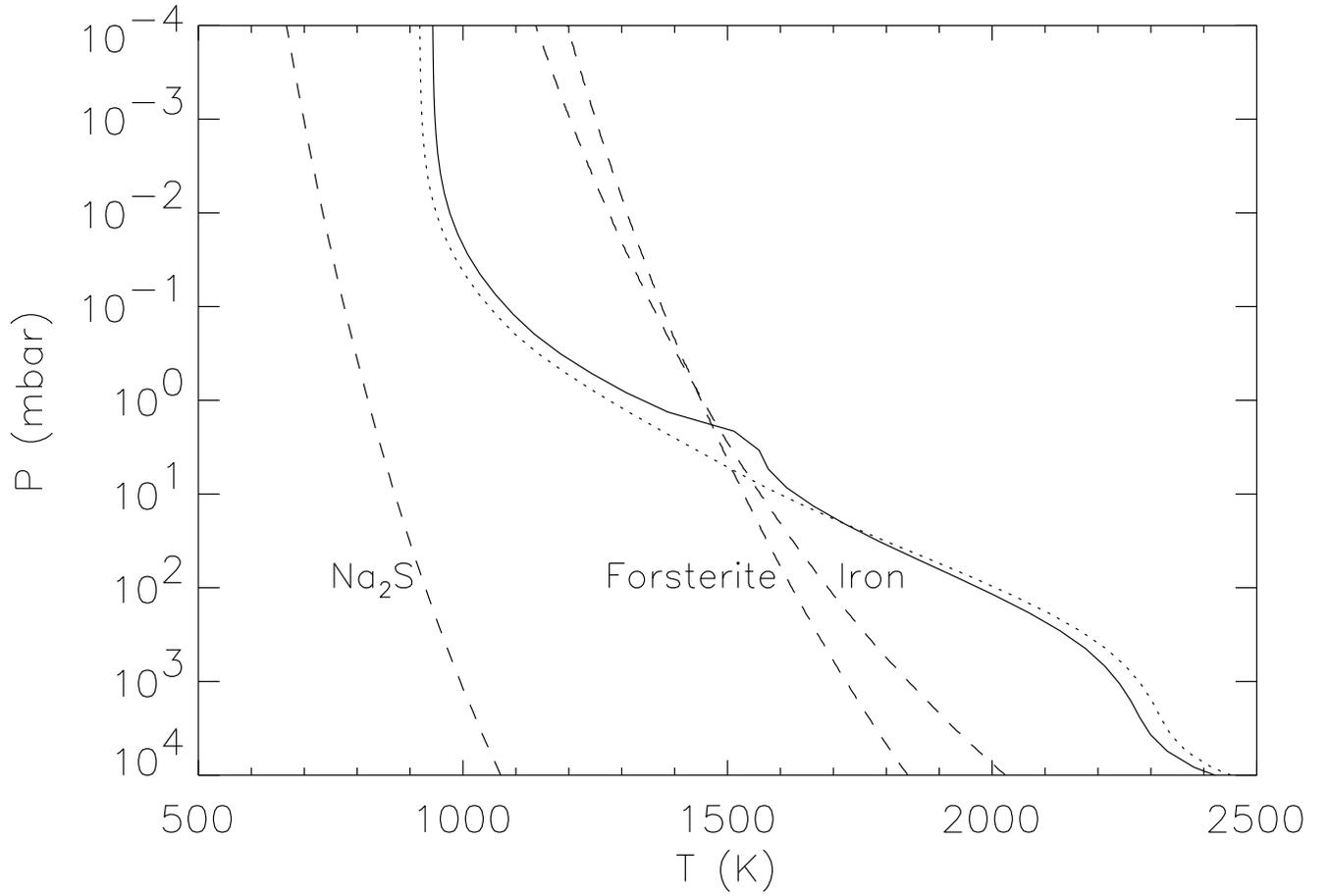}
\caption{Temperature-pressure profile for the atmosphere of HD 209458b 
(solid line). The dotted line is the \emph{T-P} profile we derive without the opacity due to 
clouds.  The condensation 
curves for Na$_2$S,
forsterite (Mg$_2$SiO$_4$), and iron are shown as dashed lines.  When cloud opacity is taken into account the forsterite cloud base moves from 5.4 mbar to 1.8 mbar.  There is a wide range of possible \emph{T-P} profiles for which forsterite and iron clouds form high in the planet's atmosphere 
but Na remains in atomic form.  We used a surface gravity of 980 cm/s$^2$ for the 
planet. 
\label{figure:pt}}
\end{figure}

\begin{figure}
\plotone{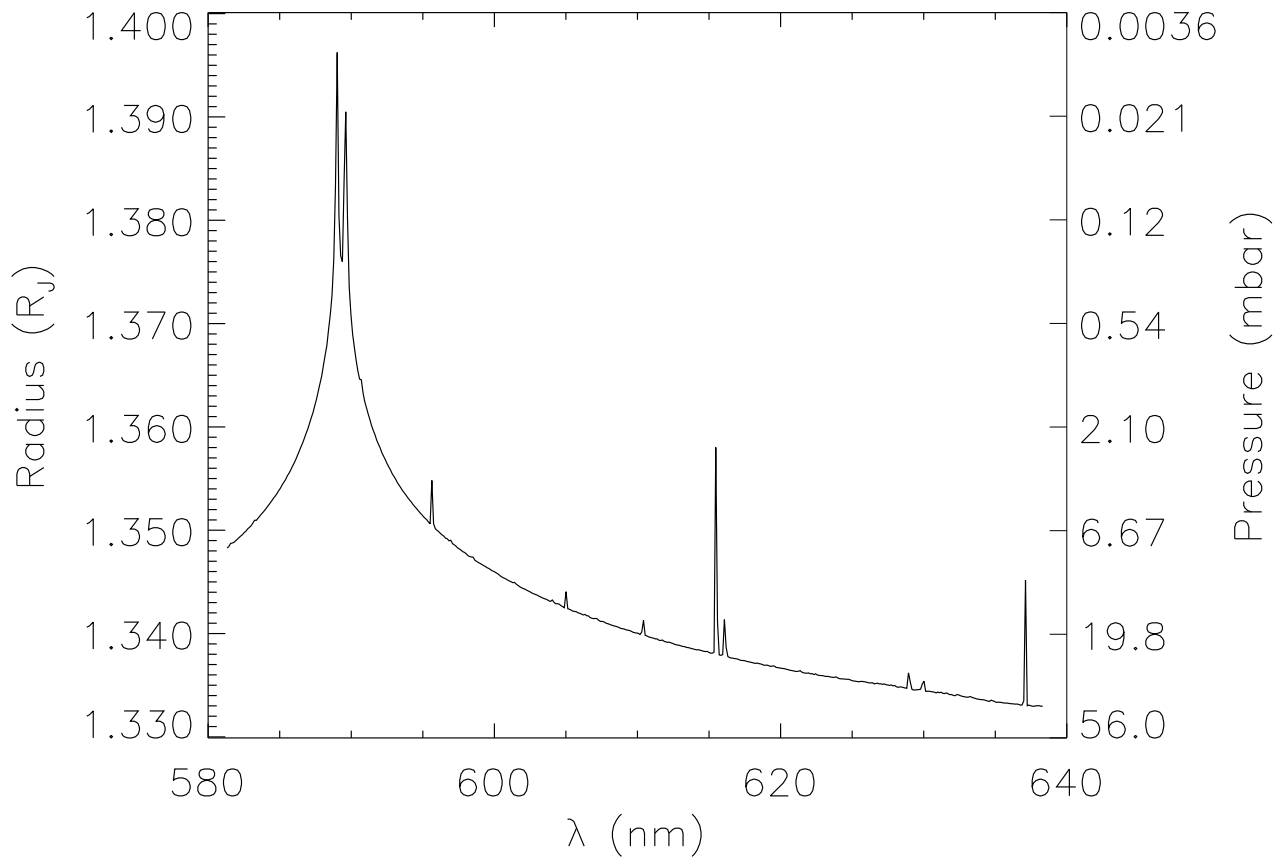}
\caption{Planetary radius versus wavelength, also showing the pressure 
at the planet's limb
corresponding to the derived radius, for a neutral cloudless atmosphere.  (1 R$_J$ = 71492 km, 
Jupiter's equatorial radius at 1 bar.) The radius values are the radii derived from calculating 
the apparent cross-sectional area of the spherical planet, and this radius is essentially 
identical to the radius 
at a slant 
opactical depth of 2/3.  The wavelengths shown encompass 
the \citet{2001ApJ...552..699B}
observations.  During the transit, an observer is sensitive to pressures from tens of mbar to
$\sim$10 $\mu$bar.  The average pressure across this wavelength band is 15 mbar.
\label{figure:pvsl}}
\end{figure}

\begin{figure}
\plotone{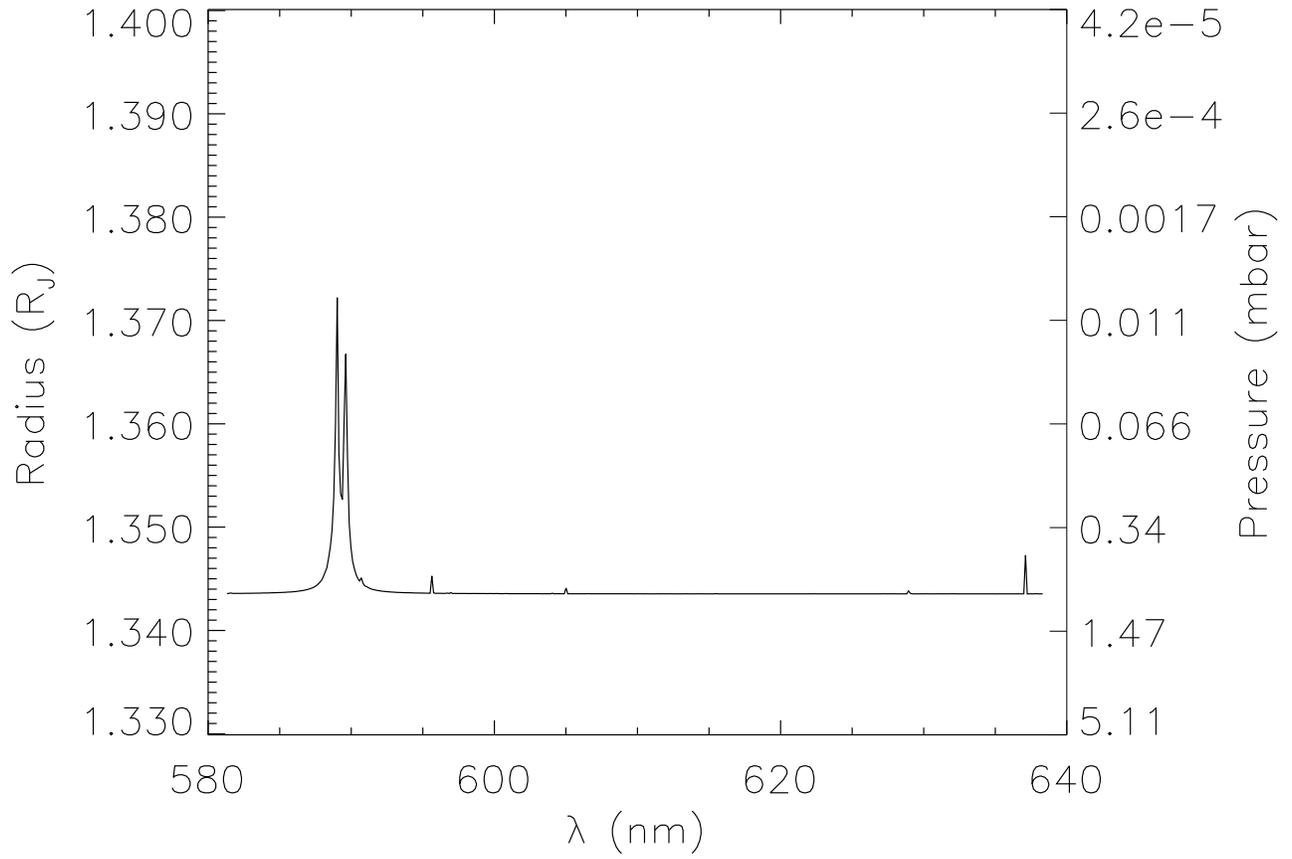}
\caption{Similar to Fig.~\ref{figure:pvsl}, but for a neutral 
atmosphere with a forsterite cloud base of 5.4 mbar.  The pressure axis changes due to the different opacity in the planet's atmosphere, as compared to Fig.~\ref{figure:pvsl}.  The clouds lead to a wavelength-independent 
increase 
in planetary radius.  The average pressure across this wavelength band is 1 mbar.
\label{figure:pvsl2}}
\end{figure}

\begin{figure}
\plotone{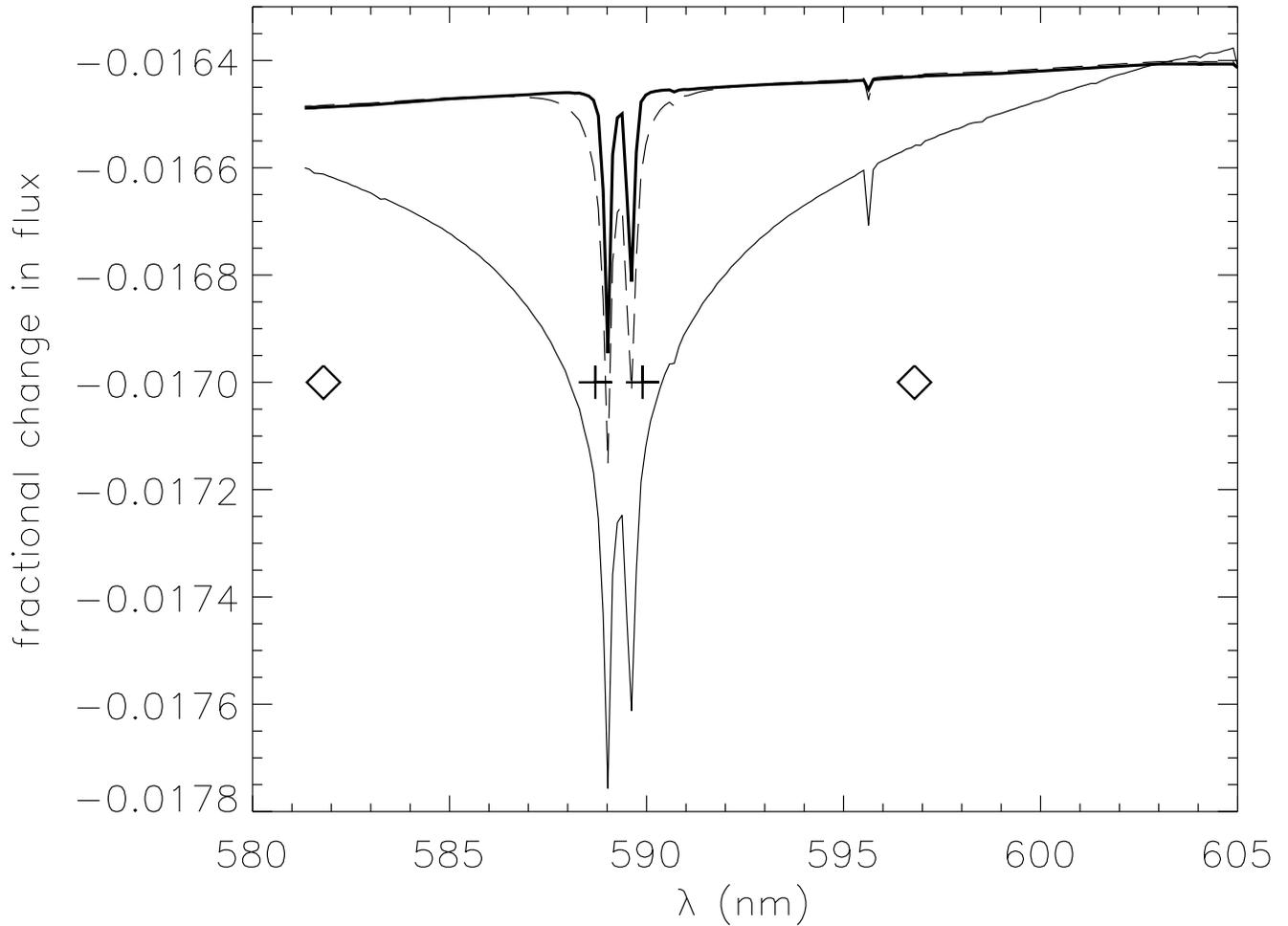}
\caption{ Transit depth (at midtransit) versus wavelength for several models.  Curves 
appear offset because the transit depth is required to
be a fractional change in flux of -0.0164 from 581 to 638 nm, the full \citet{2001ApJ...552..699B} 
observation band.  The two black plusses mark the wavelength extent of the
``in feature" band from
\citet{2002ApJ...568..377C},
while the wavelengths between the plusses and diamonds, on either side 
of the ``in feature"
band, make up the ``out of feature" band.  The observed difference in 
transit depth between these bands is $(2.32 \pm 0.57)
\times 10^{-4}$.  Shown are neutral models with and without clouds.  The Na 
feature depth for neutral cloudless model (thin solid line) is $5.20 \times 10^{-4}$.  For a 
forsterite cloud base at 5.4 mbar, the feature depth is $2.42 \times 10^{-4}$ (dashed line), and 
$0.93 \times 
10^{-4}$ for a cloud base at 1.8 mbar (thick black line).
\label{figure:dvsl2}}
\end{figure}

\begin{figure}
\plotone{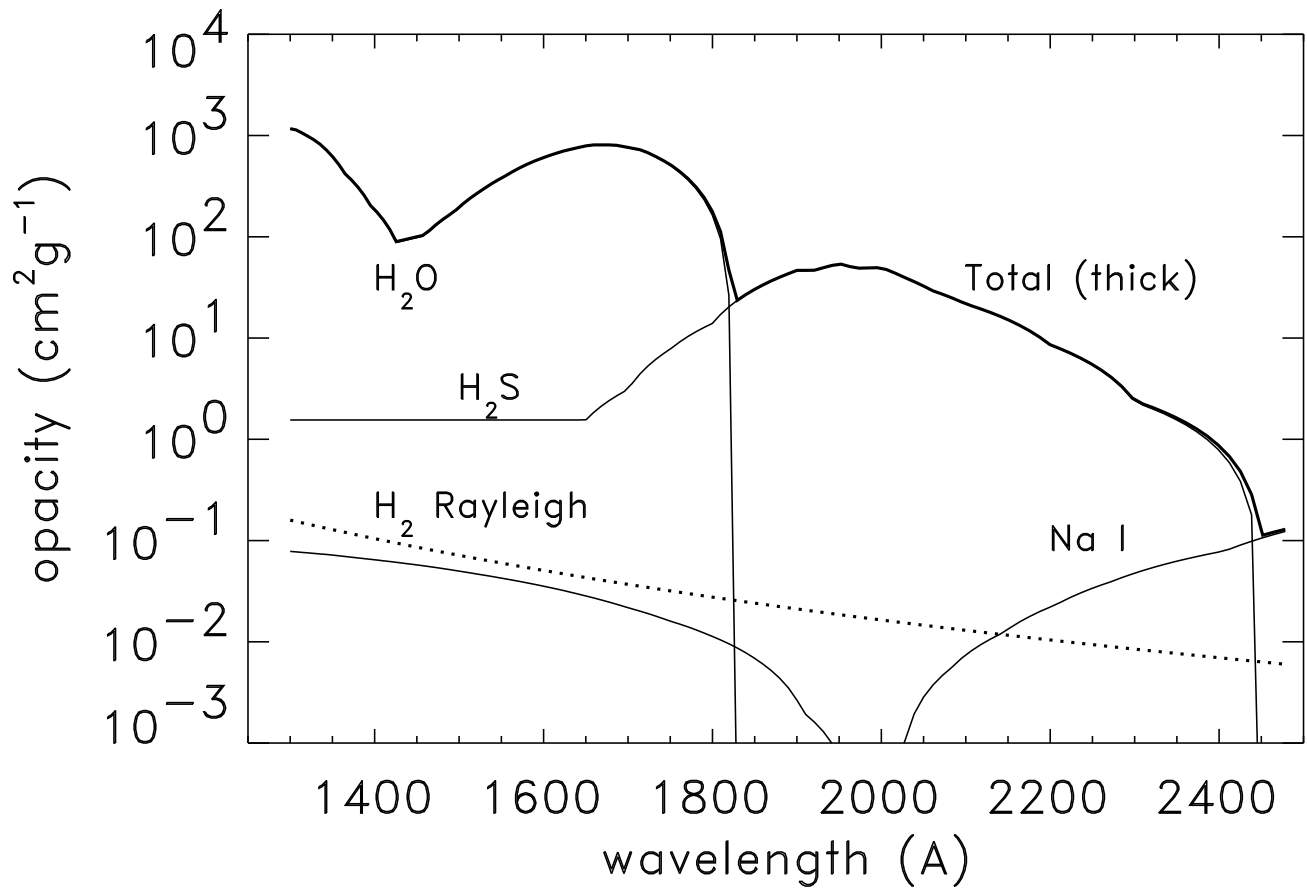}
\caption{Opacity at the outer layers of the atmosphere
of \hd.
\label{figure:opac}}
\end{figure}

\begin{figure}
\plotone{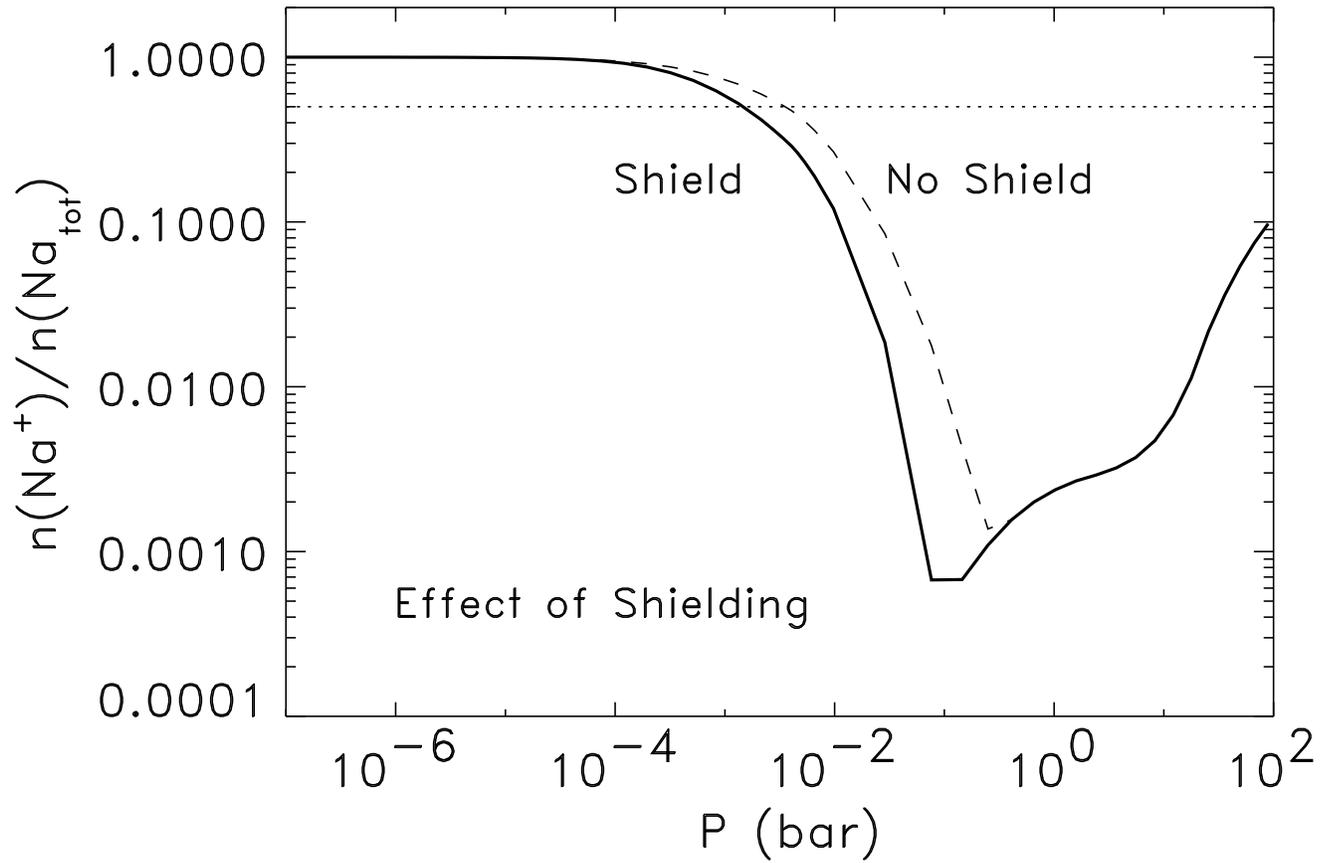}
\caption{Influence of shielding on the Na ionization depth due to UV 
absorption by H$_2$O, H$_2$S, and Rayleigh scattering.  The thick solid line is 
our nominal model, while the dashed line is the same as the nominal model, but 
with Na able to absorb $all$ ionizing radiation.  The thin horizontal dotted 
line shows the location of 50\% Na ionization.
\label{figure:shield}}
\end{figure}

\begin{figure}
\plotone{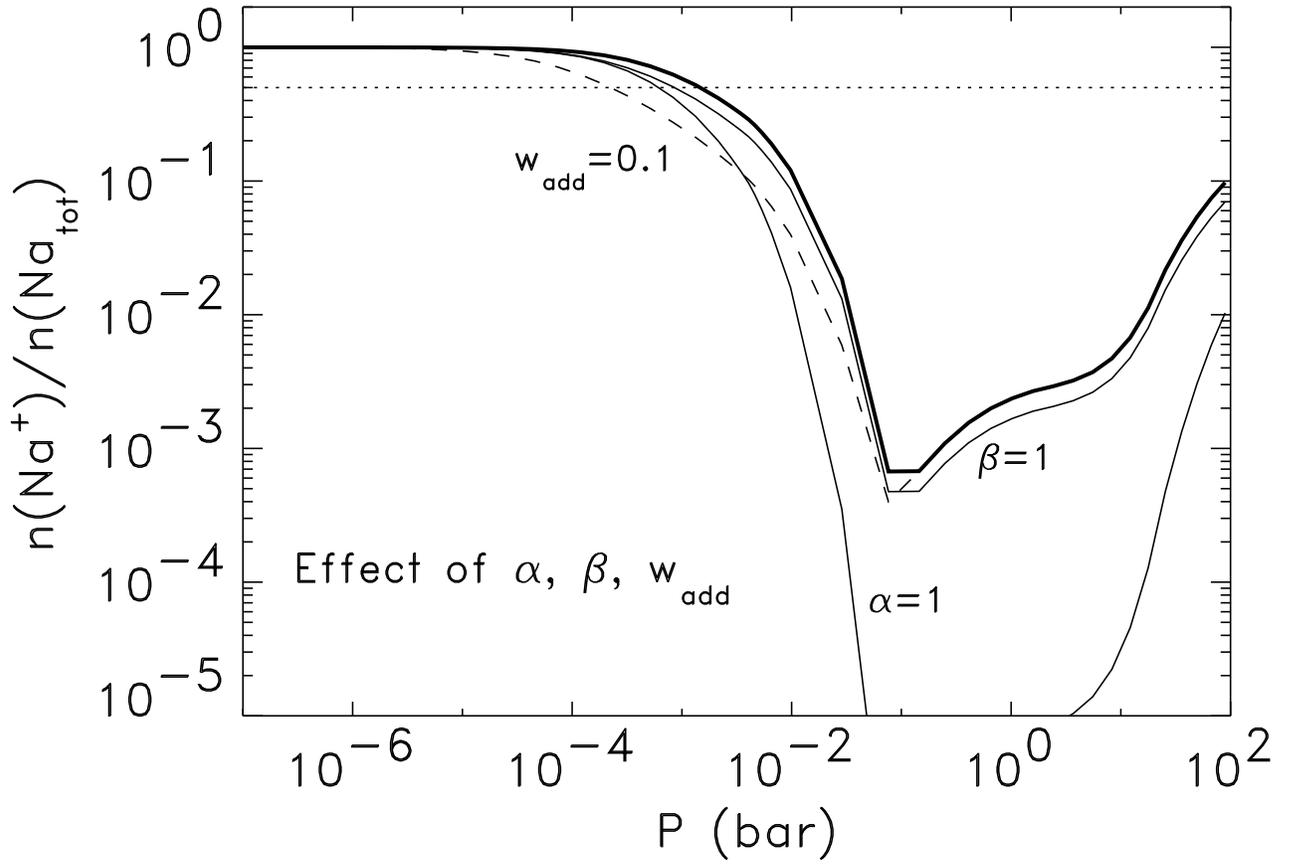}
\caption{Influence on the ionization depth of factors $\alpha$, $\beta$, 
$w_{\rm add}$, leaving other parameters at their default values.  As described 
in the text, these factors are meant to show the sensitivity of our results to 
possible unknown additional free electrons or additional UV opacity.  The thick 
solid curve is our default calculated ionization depth curve.  
($\alpha=\beta=e_{\rm H2}=0, w_{\rm add}=1$)  The thin solid curves are for 
$\alpha=1$ and $\beta=1$, while the dashed curve is $w_{\rm add}=0.1$.  The 
thin horizontal dotted line shows the location of 50\% Na ionization.  All these factors would push 
the ionization depth
out to lower pressures.
\label{figure:abw}}
\end{figure}

\begin{figure}
\plotone{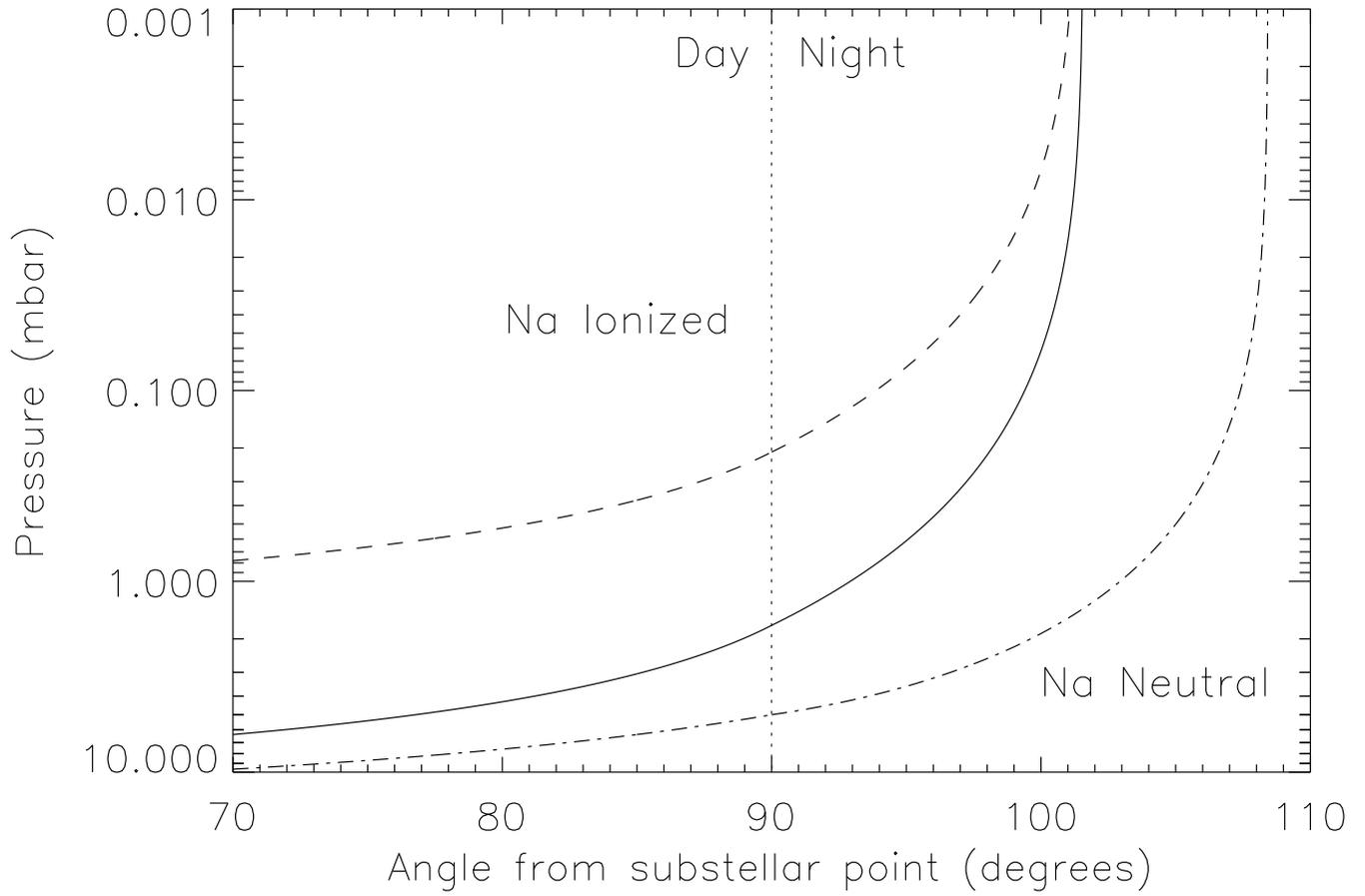}
\caption{Sodium ionization depth versus angle from the planet's
substellar point.  The dotted vertical line at 90 degrees indicates the
terminator.  The dashed curve is for model 1, while the solid curve is for model 2.  All Na is 
ionized
to the left of this curve and neutral to the right of
the curve.  The
dot-dash curve shows an increased ionization depth
at every angle, an example of an ionization depth curve that would be
necessary to match the observations in the absence of any cloud opacity.
\label{figure:pvsa}}
\end{figure}

\begin{figure}
\plotone{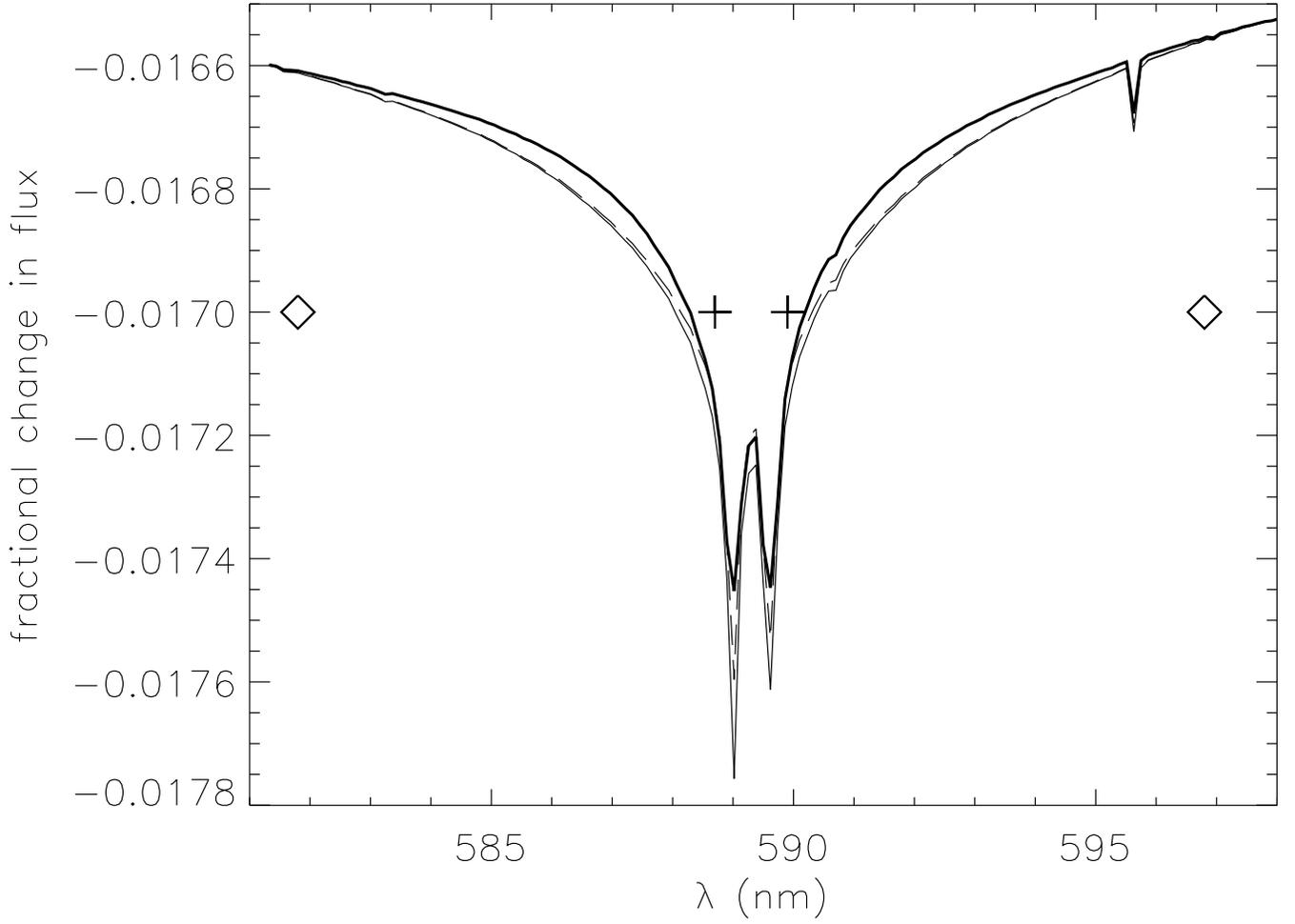}
\caption{Transit depth (at midtransit) versus wavelength for several models, similar to Fig.~\ref{figure:dvsl2}.  
The Na feature depth for neutral cloudless model (thin solid line) is $5.20 \times 
10^{-4}$.  For the ionized cases, the feature depth is $4.72 \times 10^{-4}$ for model 1 (dashed 
line), and $4.83 \times 10^{-4}$ for model 2 (thick 
solid line), respectively.  Even though the ionization depth is greater for model 2 (see Fig.~\ref{figure:pvsa} and the text), the ``in" - ``out" feature difference is greater, because in model 
2, the Na wings are affected, but in model 1 they are not. \label{figure:dvsl}}
\end{figure}

We find that HD 209458b's atmosphere is cool enough for
most heavy elements to have been incorporated into condensates but 
still warm enough
for Na to remain in atomic form.  As seen in Fig.~\ref{figure:pt}, there is 
a range of $\sim$600 K in the mbar region of the planet's 
atmosphere where this situation holds.  High clouds arise naturally in the planet's atmosphere 
near $\sim$1-5 mbar.  Although the abundances of condensable vapor of the chief cloud-forming 
species are small, the particles that form in this radiative region
 are small and have large scattering cross-sections.  Clouds lead to an increase in the apparent 
planet radius outside of the Na-D core, and on their own they could easily explain the 
observations.  In addition, clouds can be so high that the predicted 
sodium feature is $less$ than observed.

With our simple non-LTE model, we have shown that stellar UV flux will ionize Na to 
$\sim$1/2 mbar on the limb in the planet's
atmosphere, reducing the strength of the 589-nm
absorption feature.  The change in the Na-D feature due to ionization is fairly insensitive to the 
derived \emph{T-P} profile.  Ionization is a secondary effect, and it alone cannot explain the weak 
Na-D features.  This conclusion was also reached by \citet{2002ApJ...568..377C} in their analysis.  Since the majority of Na on the planet's limb at pressures $\lesssim$1/2 mbar is ionized by stellar UV flux, the possible effects of neutral Na non-LTE level populations found by \citet{2002ApJ...569L..51B} are reduced.  In addition, since the \emph{T-P} profile we derive is cooler, a region where Na LTE level populations may not hold will be pushed out to lower pressure.

For our \emph{T-P} profile, we find an opaque forsterite cloud base at a pressure 
of several mbar.  Predicting the exact location of the cloud base and vertical distribution of 
cloud material is
challenging, but the Na-D feature is quite senstive to these factors.  An 
atmosphere incorporating the forsterite cloud at 1.8 mbar and ionization gives a Na-D feature 
less than observed (due to clouds, not ionization), but for a cloud base of at 5.4 mbar, we obtain 
a feature depth that matches observations with or without ionization.

More advanced modeling of photochemistry may reveal other 
non-equilibrium
compounds, perhaps including a photochemical haze, which
could both increase the importance of particulate absorption and 
decrease the
extent to which the atmosphere is ionized. The potassium feature at 770 nm 
should be more sensitive
to ionization than is Na, because the former's ionization energy is 0.8 
eV less than that of Na.  We are currently updating our UV opacity 
database and collecting all  
available collsional parameters.  Once this is completed, it 
will be worthwhile to 
revisit the sodium ionization
problem, taking into account many levels of Na I, and computing
consistent
non-LTE level populations of sodium together with the radiation 
field.
More 
advanced global circulation 
models for \hd\ will be of great interest, as they will give a better 
indication of the temperatures expected on the limb of the 
planet.  It is likely that observations will push modeling in 
new directions, and data obtained
across many wavelengths, both from the ground and from space, including 
predicted K, H$_2$O, and CO 
features,
will help gauge the importance of processes at work in
HD 209458b's atmosphere and decide between explanations
of the Na-D feature depth.

\acknowledgments
The authors acknowledge the valuable insight gained through discussions 
with Don
Hunten and John Plane and support from grants NAG5-10760, -10450, and 
-10629.

%\newpage

\newpage

\newpage

\newpage

\newpage

\newpage 

\newpage

\newpage

\newpage

\newpage

\newpage

\end{document}